\documentstyle[epsf]{aipproc}
\begin{document}
\title{Dipenguin-like contributions to $D^0-\bar D^0$ mixing}
\author{Alexey A. Petrov}
\address{Department of Physics and Astronomy \\
The Johns Hopkins University \\
Baltimore MD 21218}
\vskip -0.3in 
\begin{flushright}{JHU-TIPAC--97021}\end{flushright}
\vskip -0.3in
\maketitle
\begin{abstract}
We show that because of the milder GIM suppression, the 
four-quark dipenguin operator contributes to the short distance
piece of the $D^0-\bar D^0$ mixing amplitude at the same order of 
magnitude as the box diagram. In addition, we put an upper bound 
$\Delta m_D^{LDR} \leq 5 \times \Delta m_D^{box}$ on a long-distance 
resonant contribution induced by the penguin mixing. 
\end{abstract}

The phenomenon of meson mixing has been studied for a long time.
Observed in the $K^0-\bar {K^0}$ and $B^0-\bar {B^0}$
systems, it provides an 
extremely sensitive test of the Standard Model (SM). 
Extremely large value of the top quark mass makes
the short distance box diagram (e.g. Fig.1a) dominate $B^0-\bar {B^0}$ 
and $K^0-\bar {K^0}$ mixing amplitudes. This fact
effectively turns GIM cancellation mechanism (vanishing of the
total $\Delta Q =2$ contribution in the limit of equal internal
quark masses) into the "GIM-enhancement" of the short-distance 
contribution to $\Delta m_P$ for $P=\{K, B\}$. Long-distance contributions
associated with the propagation of the hadronic degrees of freedom and
higher-order QCD effects are relatively unimportant for 
the mixing of $B$ and $K$ mesons. 

The situation is drastically different in the case of the
$D^0-\bar {D^0}$ system: the only available
heavy $b$-quark contribution to the box 
diagram providing $\Delta C = 2$  transition is
greatly reduced by a tiny $V_{ub}$ $CKM$ matrix element and can be
safely neglected in the analysis of the box diagram 
contribution to $\Delta m_D$ (the inclusion of the $b$-quark 
further decreases the box diagram contribution \cite{datta}).
This forces the vanishing of the effect in the SU(3) limit and implies
that only the light quark mass difference guarantees
that mixing does take place.
Calculating a box diagram and constructing the
effective Hamiltonian, we realize that the smallness of the short
distance piece is guaranteed by a factor of 
$(m_s^2-m_d^2)^2/m_c^2$ \cite{disper}
(we define $L=(1+\gamma_5)/2$)
\begin{eqnarray} \label{ham}
{\cal H}_{eff} = \frac{G_F^2}{2 \pi^2} 
\xi_s \xi_d \frac{(m_s^2-m_d^2)^2}{m_c^2}
\Bigl [ \bar u \gamma_\mu L c
 \bar u \gamma_\mu L c + 2~  \bar u R c \bar 
u R c \Bigr ]
\end{eqnarray}
with $\xi_i = V^*_{ic} V_{iu}$. We denote the operators,
entering Eq. (\ref{ham}) as $O_1$ and $O_2$ respectively.
This leads to the following expression for the 
$\Delta m_D^{box}$ \cite{datta,disper}
\begin{equation} \label{msd}
\Delta m_D^{box}  =\frac{4 G_F^2}{3 \pi^2} \xi_s \xi_d 
\frac{(m_s^2-m_d^2)^2}{m_c^2} f_D^2 m_D 
(B_D - 2 B_D')
\approx 0.5 \cdot 10^{-17} ~GeV
\end{equation}
with $f_\pi \simeq 132$ MeV, $f_D \simeq 165$ MeV, $m_s = 0.2~GeV$, and 
$B_D=B_D'=1$ in the usual vacuum saturation approximation to
\begin{equation} 
\langle D^0 | O_1 | \bar D^0 \rangle =
\frac{2}{3} \frac{f_D^2 m_D^2}{2 m_D} B_D,~~~~
\langle D^0 | O_2 | \bar D^0 \rangle =
-\frac{5}{3} \frac{m_D^2}{4 m_c^2} 
\frac{f_D^2 m_D^2}{2 m_D} B_D' .
\end{equation}
The smallness of the result of Eq.(\ref{msd}) 
triggered a set of estimates of the long-distance effects
and New Physics contributions that can 
dominate the $\Delta m_D$.

The box diagram, however, does not exhaust the set of short-distance SM
contributions. Higher order QCD corrections might significantly
modify the leading order prediction provided that they are
not proportional to the small factor $(m_s^2-m_d^2)^2/m_c^2$.
For instance, a set of diagrams (topologically
distinct from the box diagram) generated by the so-called
``dipenguin'' operators (Fig.1b.) contribute to the $D$-meson mass difference 
at the same order of magnitude as the usual box diagram \cite{ap}.
To see this, let us write an effective operator relevant to dipenguin 
$\Delta C = 2$ transition. It can be obtained from the usual $\Delta C=1$ 
penguin vertex. Neglecting a tiny dipole contribution, 
\begin{equation} \label{vertex}
\Gamma =
-\frac{G_F}{\sqrt 2} 
\frac{g_s}{4 \pi^2} F_1  
\bar u \gamma_\mu L t^a c
(g^{\mu \nu} \partial^2 - \partial^\mu \partial^\nu )
A^a_\nu.
\end{equation}
Here $t^a=\lambda^a/2$, and $F_1$ is a modified Inami-Lim (IL) function. 
Using unitarity of the CKM matrix it reads 
$F_1=\sum_i \xi_i F^i_1=\xi_s (F^s_1-F^d_1) +
\xi_b (F^b_1-F^d_1)$. It is common to 
discard the $b$-quark contribution to $F_1$ as being suppressed
by small $V_{ub}$ factors. Note, however, that it {\em enhances} 
$F_1$ by approximately $20-30 \%$. From (\ref{vertex}) we obtain the 
effective Hamiltonian
\begin{eqnarray} \label{dp}
{\cal H}_{dp}= 
- \frac{G_F^2}{8 \pi^2} 
\frac{\alpha_s}{\pi} F_1^2  
\Bigl [
(\bar u \gamma_\mu L t^a c) 
\partial^\mu \partial^\nu
(\bar u \gamma_\nu L t^a c) 
- (\bar u \gamma_\mu L t^a c)
\Box
(\bar u \gamma_\mu L t^a c)
\Bigr ]
\end{eqnarray}
In what follows, we denote two operators entering
(\ref{dp}) as $\tilde O_1$ and $\tilde O_2$.
Using the equations of motion we obtain for the first operator
in (\ref{dp})
\begin{eqnarray}
\tilde O_1 = \bar u \gamma_\mu L t^a c 
~\partial^\mu \partial^\nu~
\bar u\gamma_\nu L t^a c 
\simeq  m_c^2 ~\bar u R t^a c ~
\bar u R t^a c
\end{eqnarray}

\begin{figure}
\centerline{
\epsfbox{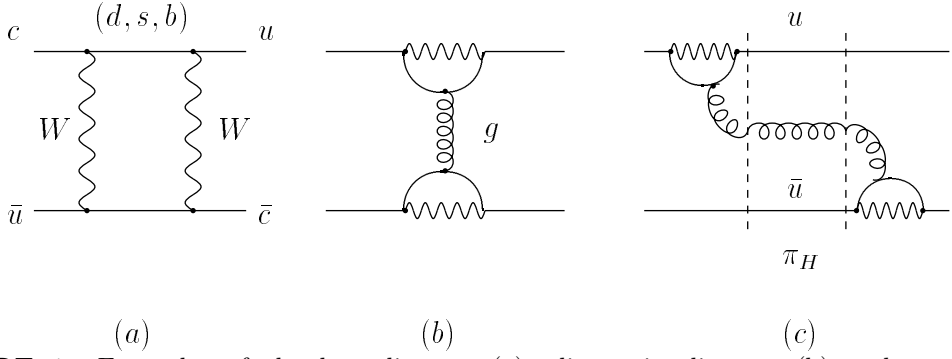}}
\caption{Examples of the box diagram (a), dipenguin diagram (b) 
and a possible
long-distance one-particle intermediate state contribution (c).}
\end{figure}

\noindent
Employing vacuum saturation method to estimate 
matrix elements and assuming that each derivative acting on the 
quark field involves an average momentum of the quark
we obtain
\begin{equation}
\langle D^0 | \tilde O_1 | \bar D^0 \rangle =
\frac{1}{9} \frac{f_D^2 m_D^4}{2 m_D} B , ~~~~~
\langle D^0 | \tilde O_2 | \bar D^0 \rangle =   
-\frac{2}{9} \frac{f_D^2 m_D^2 (2 m_c^2 - m_D^2)}{2 m_D} B'
\end{equation}
with $B$ and $B'$ being the bag parameters. This yields (setting $B=B'=1$)
\begin{equation} \label{mdp}
\Delta m_D^{dp}=2 
\langle D^0 | {\cal H}^{dp} | \bar D^0 \rangle =
\frac{G_F^2}{72 \pi^2} \frac{\alpha_s}{\pi} 
F_1^2 (m_b^2,m_s^2,m_d^2)
f_D^2 m_D (m_D^2 - 4 m_c^2).
\end{equation}
Inspecting (\ref{mdp}) we notice that contrary to $\Delta m_D^{box}$,
the dipenguin diagram does not have a power dependence upon the
internal quark masses. The leading behavior of the IL function
is logarithmic in $m_{s(d)}$, and the estimate of the 
operator brings about power dependence upon the {\em external}
quark masses, i.e.$m_c$. This feature distinguishes this contribution 
from that of the usual box diagram.
Recalling the fact that the dominant contribution to 
$K$ and $B$ mixing amplitudes is proportional to the
square of the top quark mass, it is not surprising that
this effect is small in the K and B sectors.
It is the fact of the reduced internal heavy quark dependence of the
amplitude of $D^0-\bar {D^0}$ mixing which makes the
dipenguin operator contribution effectively enhanced.
Taking into account the momentum flow $Q^2 \simeq - m_c^2$
through the gluon propagator (which is defined by the external quark 
momenta(masses)) one can calculate $F_1^i(m_i^2, Q^2) = -4 \int_0^1
dx~ x(1-x) \ln \Bigl[(m_i^2-Q^2 x (1-x))/ M_W^2 \Bigr] $
Comparing (\ref{msd}) and (\ref{mdp}) we find \cite{ap}
\begin{equation}
r = \left | \frac{\Delta m_D^{dp}}{\Delta m_D^{box}} \right | 
\approx 
\frac{\alpha_s}{8 \pi} \frac{\left[ F_1 (m_b^2,m_s^2,m_d^2)\right ]^2 
m_D^4}{| \xi_s \xi_d| (m_s^2-m_d^2)^2} 
\end{equation}
where we have put $m_c \approx m_D$, and $\alpha_s \simeq 0.4$.
The relative size of the box and dipenguin contribution 
shows that the latter is of the same order of magnitude 
as the box diagram -- our estimate
gives $r \sim 20-80 \%$ depending on the choice of  
quark masses ($m_s=0.1\div0.3~GeV$). 

The dipenguin diagram gives rise to a whole family of the 
long-distance contributions (e.g. Fig. 1c) that might be potentially 
important for $D^0-\bar D^0$ mixing.
In particular, it is plausible to assume that a gluon produced by a
penguin (or any other operator) can form an intermediate 
bound state with two up-quarks, thus producing a hybrid resonance \cite{gp}. 
These one-particle intermediate state contributions to $\Delta m_D$
might be of considerable importance because of the closure of mass of
hybrid resonance to the mass of $D$-meson. It is possible to show that
a resonant long-distance contribution to $\Delta m_D$ is given by
($D_{^L_S} = 2^{-1/2} ~\bigl [D^0 \pm \bar D^0 \bigr ]$)
\begin{equation} \label{ldm}
\Delta m_D = Re~\frac{1}{2 m_D} \sum_{n}
\frac{\left | \langle D_L | H_w | n \rangle \right |^2-
\left | \langle D_S | H_w | n \rangle \right |^2}{m_D^2-m_n^2 -
\Gamma^2/4 + i \Gamma_n m_D},
\end{equation}
where we have neglected $CP$ violation. 
The existence of the $\pi_H$ or $\pi (1800)~0^{-+}$ 
meson was reported by several experimental groups \cite{exp} and predicted 
by the theoretical calculations in various quark models \cite{theor} 
suggesting
a hybrid nature of this particle. One can estimate a contribution of
$n = \pi_H$ to Eq.(\ref{ldm}) provided that the mixing amplitude
$g = \langle D_L | H_w | \pi_H \rangle$ is known.
This amplitude can be estimated in various quark models, or
phenomenologically, by using the available data on
$D$ decay rates. The idea is to seek for the common decay channels of
$D$ and $\pi_H$ where $\pi_H$ contribution is manifest 
and then estimate the mixing amplitude using the simple model for the
$D-\pi_H$ mixing followed by $\pi_H$ decay. It was noted in 
theoretical calculations \cite{theor} and hinted experimentally that
the decay rates $\pi_H \to \pi f_0(980)$ and 
$\pi_H \to \pi f_0(1300)$ are large for the hybrid $\pi_H$. 
Thus one can put an upper bound on the 
mixing amplitude $g$ by introducing a model for the resonant 
decay of $D$-meson via $\pi_H$:
\begin{equation}
{\cal M} (D \to \pi f_0(980)) = g~(m_D^2 - m_{\pi_H}^2 + 
i \Gamma_{\pi_H} m_D)^{-1} ~A(\pi_H \to \pi f_0(980)).
\end{equation}
Computing the mixing amplitude $g$ using experimental data on the decay rate 
$\Gamma(D \to \pi f_0(980))$ and inserting it into Eq. (\ref{ldm}), 
we estimate $\Delta m_D^{\pi_H} \leq 0.2 \times 10^{-16}~GeV$, which is five times 
larger than the short distance result. 

In conclusion, we have shown that the short-distance dipenguin diagram
contributes at the same order of magnitude as the box diagram to 
$D^0-\bar {D^0}$ mixing amplitude. Long-distance dipenguin-like pieces 
presumably give larger contributions, but require further studies \cite{disper,gp}.


\begin{references}
\bibitem{datta}{A. Datta and D. Kumbhakar, 
Z. Phys. C {\bf 27}, 515 (1985); H.Y. Cheng, Phys. Rev. 
D{\bf 26}, 143 (1982)}

\bibitem{disper}{J. Donoghue, E. Golowich, B. Holstein, and 
J. Trampetic, Phys. Rev. D {\bf 33}, 179 (1986);
L. Wolfenstein, Phys. Lett. B {\bf 164}, 170 (1985);
H. Georgi, {\it ibid.} {\bf 297}, 353 (1992);
T. Ohl, G. Ricciardi, and E. Simmons, Nucl. Phys.
{\bf B403}, 605 (1993).}

\bibitem{ap}{A.A. Petrov, Phys. Rev. {\bf D56}, 1685 (1997).}

\bibitem{exp}{D.V. Amelin, et.al. Phys. Lett. {\bf 356} (1995) 595.}

\bibitem{theor}{F. Close, P. Page Nucl. Phys. {\bf B443} (1995) 233;
Phys. Rev. {\bf D52} (1995) 1706.}

\bibitem{gp}{A. Zaitsev, private communications;  
E. Golowich and A.A. Petrov, in preparation}

\end{references}
\end{document}